\documentclass[lettersize,journal]{IEEEtran}
\usepackage{amsmath,amsfonts}
\usepackage{mathrsfs}
\usepackage[ruled,linesnumbered]{algorithm2e}
\usepackage{array}
\usepackage[caption=false,font=normalsize,labelfont=sf,textfont=sf]{subfig}
\usepackage{textcomp}
\usepackage{stfloats}
\usepackage{url}
\usepackage{verbatim}
\usepackage{graphicx}
\usepackage{cite}
\usepackage{amssymb}
\usepackage{cuted}
\usepackage{color}
\usepackage{eurosym,booktabs,multirow}
\usepackage{verbatim}
\usepackage{cite}
\usepackage{makecell}
\usepackage{subfig}
\usepackage{stfloats}
\usepackage{enumitem}
\usepackage[dvipsnames]{xcolor}

\begin{document}

\title{\LARGE A Heuristic-Integrated DRL Approach for Phase Optimization in Large-Scale RISs}
\author{
Wei~Wang,~\IEEEmembership{Graduate Student Member,~IEEE}, Peizheng~Li,~\IEEEmembership{Member,~IEEE}, Angela~Doufexi,~\IEEEmembership{Member,~IEEE}, and Mark~A.~Beach,~\IEEEmembership{Senior Member,~IEEE}
\thanks{This work was supported by China Scholarship Council (CSC) under grant 202108060224.}
\thanks{Wei~Wang, Angela~Doufexi, and Mark~A.~Beach are with the Communication Systems and Networks Research Group, University of Bristol, Bristol BS8 1UB, U.K.
(e-mail:~wei.wang@bristol.ac.uk; a.doufexi@bristol.ac.uk; and m.a.beach@bristol.ac.uk).
}
\thanks{Peizheng~Li is with the Bristol Research \& Innovation Laboratory, Toshiba Europe Ltd., Bristol BS1 4ND, U.K. (e-mail:~peizheng.li@toshiba-bril.com).
}
}



\maketitle

\begin{abstract}
Optimizing discrete phase shifts in large-scale reconfigurable intelligent surfaces (RISs) is challenging due to their non-convex and non-linear nature. In this letter, we propose a heuristic-integrated deep reinforcement learning (DRL) framework that \emph{(1)} leverages accumulated actions over multiple steps in the double deep Q-network (DDQN) for RIS column-wise control and \emph{(2)} integrates a greedy algorithm (GA) into each DRL step to refine the state via fine-grained, element-wise optimization of RIS configurations. By learning from GA-included states, the proposed approach effectively addresses RIS optimization within a small DRL action space, demonstrating its capability to optimize phase-shift configurations of large-scale RISs.
\end{abstract}

\vspace{-5pt}

\section{Introduction}
\IEEEPARstart{R}{econfigurable} intelligent surface (RIS) is a promising technology for 6G networks by enabling intelligent adaptivity of the wireless propagation environment. Typically, an RIS consists of numerous unit cells, the phase shift configuration of which plays a crucial role in enhancing the system performance of RIS-assisted networks, including data rate, energy efficiency, channel capacity, etc.~\cite {9424177}. To efficiently optimize this phase shift configuration---a non-convex and non-linear optimization problem---deep reinforcement learning (DRL) has been widely studied~\cite{10500737, 10037180, 10458888, 10261216, 10684115, 9558821, 9919620, 9277627, 9526285, 10539342, 10500993}, as it provides a generic, trial-and-error-based framework that interacts with the training environment to maximize cumulative rewards.

Based on the continuous phase shift assumption of RISs, continuous DRL algorithms like deep deterministic policy gradient (DDPG), proximal policy optimization (PPO), and soft actor-critic (SAC) have been investigated~\cite{10458888, 10261216, 10684115, 9558821}. However, the practical implementation of RISs requires discrete phase shifts, while quantizing continuous configurations to discrete configurations leads to performance degradation~\cite {9424177}. Therefore, \cite{9919620, 9277627, 10539342, 9526285} considered the discrete DRL algorithms, such as double deep Q-network (DDQN)~\cite{9919620, 9277627}, Dueling DQN~\cite{9526285}, and dueling double deep Q-network (D3QN)~\cite{10539342}, to directly optimize the phase shifts of each RIS unit cell through element-wise control in a discrete parameter space. While this approach provides RISs with the highest degree of freedom (DoF) for exploring the optimal configurations, it also results in an exponentially growing exploration space as the number of RIS unit cells increases. Specifically, for an $N$-element RIS with $R_{\theta}$-bit discrete resolution, the possible configurations and thus exploring complexity expand to $2^{NR_{\theta}}$ which is too huge for discrete DRL algorithms to converge if $N$ is large~\cite{10021676}, making them impractical for large-scale RISs.

Large-scale RISs compensate for the low efficiency of passive RIS unit cells and severe path loss~\cite{9810144} but introduce significant optimization challenges. To address this, \cite{10037180} grouped the RIS unit cells in the same column into one group sharing the same phase shift and applied DQN for column-wise control with 1-bit resolution. \cite{10500737} grouped the adjacent 10 unit cells into one group and utilized a greedy algorithm (GA) as a pre-optimization to reduce exploration space for DRL training further. Although they significantly reduce the exploration space, the number of possible configurations remains exponential to the number of groups and resolutions, since all potential configurations must be included in their DRL action space.
Different from grouping, the coarse phase control method in~\cite{9794416} offered only limited actions (such as 3, 5, or 32 actions), with each action pre-defined as an incremental phase-shift vector of the current configuration. This method overcomes the exponentially increasing complexity of DRL-optimized RISs, but the bit-wise adjustments made in sequence cause RIS unit cells that appear earlier to be optimized less, while those that appear later are optimized more, resulting in imbalanced optimization.

This letter proposes an effective heuristic-integrated DRL approach to optimize discrete phase shift configurations for large-scale RISs. The contributions are summarized as follows:
\begin{itemize}[leftmargin=*]
    \item We employ DDQN for column-wise RIS control. Unlike previous works~\cite{10500737,10037180} treating all possible configurations as DRL actions, we define actions as column indices, where a fixed value increases the corresponding phase shifts. These incremental adjustments are accumulated over several steps within each episode to complete the RIS configuration.
    \item Then, we integrate a GA at each DRL step to refine the RIS unit cell control within the DDQN-selected column. By leveraging information before and after the GA's application, the DDQN is trained to select actions that offer higher potential for the following GA's element-wise optimization.
    \item As a result, the proposed algorithm significantly reduces the action space size of discrete DRL algorithms from exponential ($2^{NR_{\theta}}$) to linear square-root complexity ($\sqrt{N}$) by accumulating column-indexing actions. Meanwhile, integrating GA further compensates for the reduced optimization DoF and performance due to grouping, enabling element-wise phase shift optimization for large-scale RISs within a limited DRL action space.
\end{itemize}

\section{System Model}
\label{sec: system model}
This letter considers a multiuser downlink near-field communications network in an indoor scenario, involving an $M$-antenna femtocell base station (FBS), an $N$-element rectangular-array RIS, and $K$ single-antenna users, with their sets denoted by $\mathcal{M} \triangleq \{1,2,..., M\}$, $\mathcal{N} \triangleq \{1,2,..., N\}$, and $\mathcal{K} \triangleq \{1,2,..., K\}$, respectively.
To simplify the analysis and focus on investigating the optimization of the discrete RIS phase shift matrix, the channel state information (CSI) is assumed to be perfectly known at the FBS, and the direct links between FBS and users are assumed to be blocked. Thus, the received signal $y_k, \forall k \in \mathcal{K}$ at $k$th user can be expressed as
\begin{equation}
\setlength\abovedisplayskip{5pt}
\setlength\belowdisplayskip{4pt}
    y_{k} = (\boldsymbol{\rm{h}}_{k}^{T} \boldsymbol{\Phi}\boldsymbol{\rm{G}}) \boldsymbol{\rm{x}} + \it{n}_{\emph{k}},
\label{equi:signal_1}
\end{equation}
where the transmitted signal vector $\boldsymbol{\rm{x}}=\sum_{k=1}^{K}\sqrt{p_k} \boldsymbol{{\rm w}}_{k}s_{k} \in \mathbb{C}^{M}$, with $s_{k}$ denoting the desired signal for the $k$th user, alongside its corresponding precoding vector $\boldsymbol{{\rm w}}_{k} \in \mathbb{C}^{M}$ and allocated transmit power $p_k$. $p_k$ should meet the maximum transmit power constrain of FBS that $\sum_{k=1}^{K} p_k \leq P_{\rm max}$.
$\it{n}_{\emph{k}} \in \mathbb{C}$ denotes the received noise at $k$th user with zero mean and variance $\sigma^{2}_{k}$.
Additionally, the Rician channels $\boldsymbol{\rm{h}}_{k} \in \mathbb{C}^{N}$ between the RIS and the $k$th user, and $\boldsymbol{\rm{G}} \in \mathbb{C}^{N\times M}$ between the FBS and the RIS, can be further expressed as
\begin{equation}
\boldsymbol{\rm{h}}_{k}=\sqrt{\tfrac{\varepsilon_{h}}{\varepsilon_{h}+1}}\overline{\boldsymbol{\rm{h}}}_{k}+\sqrt{\tfrac{1}{\varepsilon_{h}+1}}\widetilde{\boldsymbol{\rm{h}}}_{k},
\label{equi:Rician_1}
\end{equation}
\begin{equation}
\setlength\abovedisplayskip{5pt}
\setlength\belowdisplayskip{4pt}
\boldsymbol{\rm{G}}=\sqrt{\tfrac{\varepsilon_{G}}{\varepsilon_{G}+1}}\overline{\boldsymbol{\rm{G}}}+\sqrt{\tfrac{1}{\varepsilon_{G}+1}}\widetilde{\boldsymbol{\rm{G}}},
\label{equi:Rician_2}
\end{equation}
where $\overline{\boldsymbol{\rm{h}}}_{k}$ and $\overline{\boldsymbol{\rm{G}}}=[\overline{\boldsymbol{\rm{g}}}_{1},\overline{\boldsymbol{\rm{g}}}_{2},...,\overline{\boldsymbol{\rm{g}}}_{M}]$ represent the line-of-sight (LoS) components while $\widetilde{\boldsymbol{\rm{h}}}_{k}$ and $\widetilde{\boldsymbol{\rm{G}}}$ denote the non-line-of-sight (NLoS) components, with the Rician factors $\varepsilon_{h}$ and $\varepsilon_{G}$ describing their relative weights in the propagation.

The RIS beamforming matrix $\boldsymbol{\Phi}$ in (\ref{equi:signal_1}) is a diagonal matrix given by
\begin{equation}
\setlength\abovedisplayskip{5pt}
\setlength\belowdisplayskip{4pt}
    \boldsymbol{\Phi} \triangleq \beta{\rm diag}(\boldsymbol{\phi})= \beta\rm{diag}({[e^{j\theta_{1}},e^{j\theta_{2}},...,e^{j\theta_{N}}]}^{\it{T}}),
\label{equi:RIS}
\end{equation}
where $\beta$ represents the amplitude response, which is assumed to be uniform across all RIS elements so that the RIS only applies the reconfigurable phase shifts $\Theta \triangleq {[\theta_{1}, \theta_{2}, ..., \theta_{N}]}^T$ to the incoming signals with $\theta_n$ denoting the discrete phase shift applied by the $n$th RIS element, modeled as
\begin{equation}
\setlength\abovedisplayskip{5pt}
\setlength\belowdisplayskip{4pt}
    \theta_n \in \mathcal{P} \triangleq \left\{j \tfrac{2\pi}{2^{R_{\theta}}} \mid j=0,1,..., 2^{R_{\theta}}-1 \right\},
\label{equi:phase}
\end{equation}
where the positive integer values $R_{\theta} \in \mathbb{Z}^+$ is the resolution of the available discrete phase shifts in bits.

By defining $\boldsymbol{\rm{h}}_{\rm{ris}\it{,k}}\triangleq \boldsymbol{\rm{h}}_{k}^{T} \boldsymbol{\Phi}\boldsymbol{\rm{G}} \in \mathbb{C}^{1\times M}$, the signal-to-interference-plus-noise ratio (SINR) of the $k$th user can be formulated as
\vspace{-3pt}
\begin{equation}
\setlength\abovedisplayskip{5pt}
\setlength\belowdisplayskip{4pt}
{\rm SINR}_{k} = \dfrac{{p_k\mid \boldsymbol{\rm{h}}_{\rm{ris}\it{,k}}~\boldsymbol{\rm{w}}_{k} \mid}^{2}}{\sum^{K}_{j=1,j\neq k}{p_j\mid \boldsymbol{\rm{h}}_{\rm{ris}\it{,k}}~\boldsymbol{\rm{w}}_{j} \mid}^{2}+\sigma^{2}_{k}}.
\label{equi:SINR_k}
\end{equation}
\vspace{-3pt}

Since we focus on the phase shift optimization of RIS in this letter, we apply the regularized zero-forcing (RZF) precoding to balance the noise suppression and interference mitigation, along with uniformly distributed power allocation ($p_k=\dfrac{P_{\rm max}}{K}$) to simplify the problem. The precoding matrix $\boldsymbol{\rm{W}} \triangleq [\boldsymbol{\rm{w}}_{1},\boldsymbol{\rm{w}}_{2},...,\boldsymbol{\rm{w}}_{K}] \in \mathbb{C}^{M\times K}$ under the RZF is given by

\begin{small}
\begin{equation}
\setlength\abovedisplayskip{5pt}
\setlength\belowdisplayskip{4pt}
\boldsymbol{\rm{W}}=\left\{
\begin{array}{cl}
\dfrac{\boldsymbol{\rm{H}}_{\rm{ris}}^{H}{(\boldsymbol{\rm{H}}_{\rm{ris}} \boldsymbol{\rm{H}}_{\rm{ris}}^{H} + \kappa \boldsymbol{\rm I}_{K})}^{-1}}{\parallel \boldsymbol{\rm{H}}_{\rm{ris}}^{H}{(\boldsymbol{\rm{H}}_{\rm{ris}} \boldsymbol{\rm{H}}_{\rm{ris}}^{H} + \kappa \boldsymbol{\rm I}_{K})}^{-1} \parallel},\ K\leq M,\vspace{1.5ex}\\
\dfrac{{(\boldsymbol{\rm{H}}_{\rm{ris}}^{H} \boldsymbol{\rm{H}}_{\rm{ris}} + \kappa \boldsymbol{\rm I}_{M})}^{-1} \boldsymbol{\rm{H}}_{\rm{ris}}^{H}}{\parallel {(\boldsymbol{\rm{H}}_{\rm{ris}}^{H} \boldsymbol{\rm{H}}_{\rm{ris}} + \kappa \boldsymbol{\rm I}_{M})}^{-1} \boldsymbol{\rm{H}}_{\rm{ris}}^{H} \parallel},\ K> M,\\
\end{array} \right.
\label{equi:precoding}
\end{equation}
\end{small}
where the cascaded channel matrix is defined as $\boldsymbol{\rm{H}}_{\rm{ris}} \triangleq {[\boldsymbol{\rm{h}}_{{\rm ris},1},\boldsymbol{\rm{h}}_{{\rm ris},2},...,\boldsymbol{\rm{h}}_{{\rm ris},K}]}^{T} \in \mathbb{C}^{K\times M}$. The regularization parameter $\kappa = K {\sigma}_{k}^{2}$ is with the assumption that all $K$ users have the same noise power~\cite{mao2022rate}. $\boldsymbol{\rm I}_{K} \in \mathbb{R}^{K\times K}$ and $\boldsymbol{\rm I}_{M} \in \mathbb{R}^{M\times M}$ are the identity matrices.

Therefore, the sum data rate of this multiuser downlink transmission, where all users share a total bandwidth of $B$ over the channel, can be expressed as a function of $\boldsymbol{\Phi}$ as
\vspace{-3pt}
\begin{equation}
\setlength\abovedisplayskip{5pt}
\setlength\belowdisplayskip{4pt}
\mathcal{R}_{\rm{sum}}({\boldsymbol {\Phi}}) = \sum_{k=1}^{K} B\,{\rm log}_{2} \left(1+{\rm SINR}_{k}({\boldsymbol {\Phi}})\right).
\label{equi:SumRate}
\vspace{-3pt}
\end{equation}

In this letter, we aim to maximize this aforementioned system sum rate by optimizing the discrete phase shift matrix for RIS, formulated as
\begin{small}
\begin{IEEEeqnarray}{ccl}\label{P_1}
\setlength\abovedisplayskip{5pt}
\setlength\belowdisplayskip{4pt}
\boldsymbol{\rm{P}}: \,\, &\underset{\boldsymbol{\rm{\Phi}}}{\text{maximize}}\ & \mathcal{R}_{\rm{sum}},
\\
	&\text{s.t.} &\theta_n \in \mathcal{P}, \forall n \in \mathcal{N}. \IEEEyessubnumber \label{P_1a}
\vspace{-5pt}
\end{IEEEeqnarray}
\end{small}

Note that this is a non-convex and non-linear integer optimization problem, where the number of potential solutions $2^{NR_{\theta}}$ grows exponentially with the number of RIS unit cells $N$ and resolution levels $R_{\theta}$. Solving this problem within a discrete parameter space for large-scale RISs is challenging for numerical optimization methods, while attempting to address it in a continuous parameter space with relaxation and quantization also leads to performance degradation. Therefore, we employ a DRL-based algorithm to solve it.
\vspace{-5pt}

\section{Proposed Heuristic-integrated DRL Optimization Algorithm}
\label{sec: optimization}
\vspace{-5pt}
In this section, we introduce a proposed novel, efficient, and effective DDQN-based algorithm for column-wise RIS optimization, which significantly reduces the action space, and we further incorporate a GA for element-wise fine optimization.
\vspace{-10pt}
\subsection{Proposed DRL Algorithm for Column-wise Phase Shift Design}
DDQN is serving as the foundation for direct RIS optimization in the discrete parameter space. 
By separating action selection and evaluation, DDQN achieves better training stability and convergence efficiency compared to its predecessor, DQN.
The details of the proposed DDQN are listed as follows:
\begin{itemize}[leftmargin=*]
    \item State: The state $s_t \in \mathbb{R}^{N\times (2M+2K+1)}$ at step $t$ includes the real parts $\Re(\boldsymbol{\rm{G}})\in \mathbb{R}^{N\times M}$ and $\Re(\boldsymbol{\rm{H}})\in \mathbb{R}^{N\times K}$ and the imaginary parts $\Im(\boldsymbol{\rm{G}})\in \mathbb{R}^{N\times M}$ and $\Im(\boldsymbol{\rm{H}})\in \mathbb{R}^{N\times K}$ of the FBS-RIS channel $\boldsymbol{\rm{G}}$ and RIS-users channel $\boldsymbol{\rm{H}}\triangleq [\boldsymbol{\rm{h}}_1,\boldsymbol{\rm{h}}_2,...,\boldsymbol{\rm{h}}_K]$, respectively, as well as the RIS phase shift in degrees $\Theta_{t-1}\in \mathbb{R}^{N}$ at step ($t-1$). Therefore, it is represented as
    \begin{equation}
    \setlength\abovedisplayskip{5pt}
    \setlength\belowdisplayskip{4pt}
        s_t = [\Re(\boldsymbol{\rm{G}}), \Re(\boldsymbol{\rm{H}}), \Im(\boldsymbol{\rm{G}}), \Im(\boldsymbol{\rm{G}}), \Theta_{t-1}].
    \end{equation}

    \item Action: The action $a_t\in \{0, 1, ..., \sqrt{N}\}$ is defined as an index that determines whether to increase the phase shift of the corresponding RIS column to the next discrete value in $\mathcal{P}$, with $a_t=0$ indicating that the RIS phase shift configuration will remain unchanged during that step.

    \item Reward: Since the optimizing object is to maximize the system sum rate, we define the reward as $r_t=\mathcal{R}_{\rm{sum}}$ for all steps except the final step in each episode, which is defined as $r_t= \omega \mathcal{R}_{\rm{sum}}$ with $\omega$ being a weight parameter.
\end{itemize}

\begin{figure}[t]
\centering
\includegraphics[width=80mm]{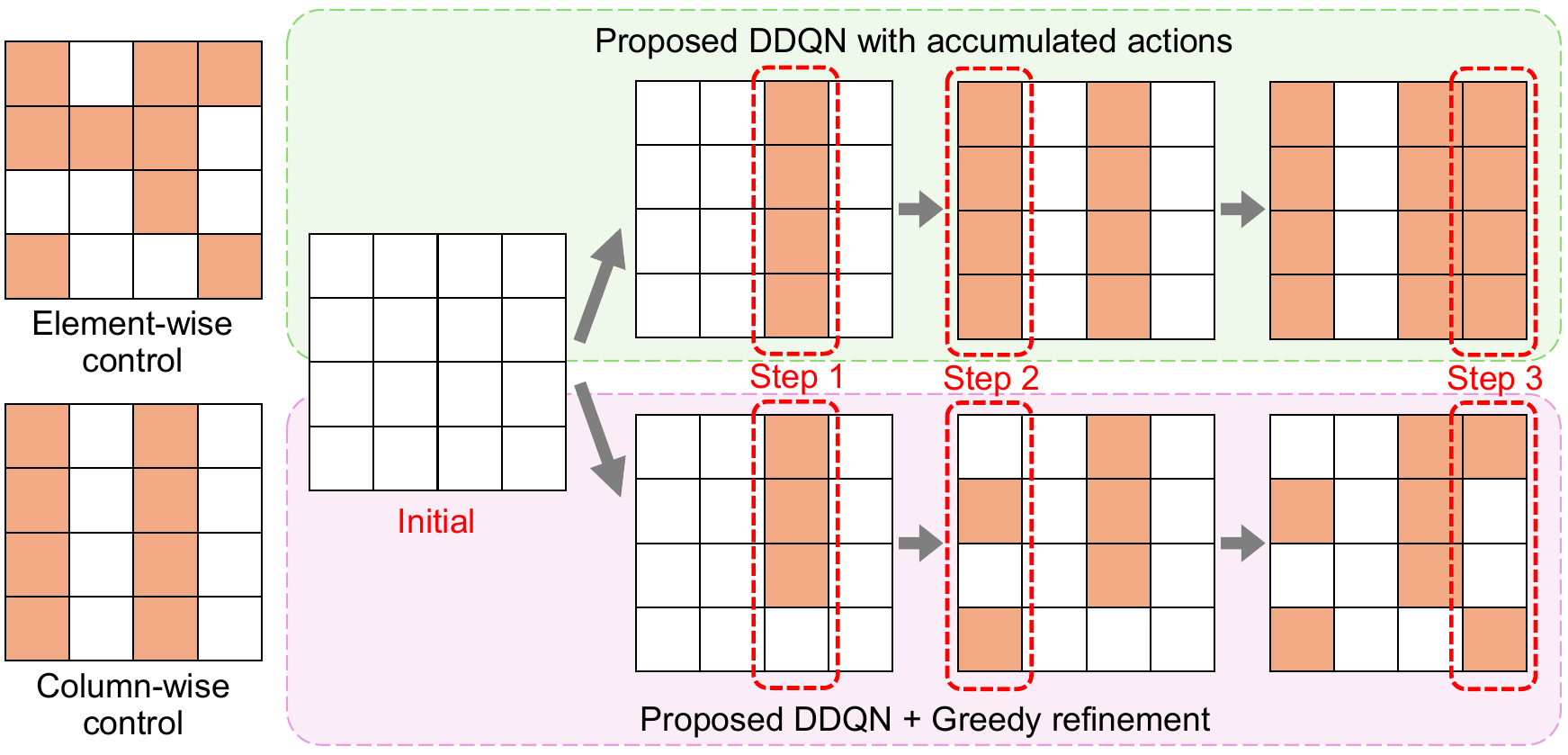}
\caption{Illustration of the proposed DDQN and DDQN-Greedy schemes.}
\vspace{-10pt}
\label{fig:DRL}
\end{figure}

Fig.~\ref{fig:DRL} illustrates the proposed schemes compared to conventional DRL methods for RIS configuration optimization. For a 1-bit $4\times4$ RIS array, element-wise controlled DRL~\cite{9526285, 9277627} requires an action space of $2^{16} = 65{,}536$, while column-wise control~\cite{10037180} reduces it to $2^4 = 16$. By adopting the proposed accumulated-action design, the action space is further reduced to 5, where $a_t = 0$ denotes no change and $a_t = {1,2,3,4}$ increases the corresponding RIS column to its next state in $\mathcal{P}$. Unlike element-wise and column-wise methods that generate a complete RIS configuration in a single step, the proposed scheme accumulates actions over multiple steps to form one configuration. As a result, the proposed DDQN-based scheme significantly reduces the DRL action space from the exponential growth of $2^{N R_{\theta}}$ to a linear square-root dependency $\sqrt{N}$, thereby facilitating effective and efficient large-scale RIS optimization.
The detailed structure of the proposed DRL algorithm is summarized in Algorithm.~\ref{Algorithm 1}, with the parameters explained in TABLE~\ref{table: DRL parameter}.
Specifically, at the beginning of each episode, the RIS phase shift configuration $\Theta_0$ is initialized as an all-zero vector. Subsequently, the agent selects an action either randomly with probability $\epsilon=\max(\epsilon_{\rm min}, \epsilon_{\rm init}\times{(1-\epsilon_{\rm decay})}^{iT+t})$, or based on the policy by choosing $a = \underset{a}{\rm arg max}~Q(s, a; \varrho)$ with probability $1-\epsilon$. In the subsequent steps within the episode, the selected actions are sequentially applied to update the RIS configuration, gradually forming the final RIS phase-shift configuration.
\vspace{-5pt} 

\begin{algorithm}[t]
\caption{Proposed DDQN Algorithm}\label{Algorithm 1}
\LinesNumbered
{\bf Input}: $\gamma$, $\alpha$, $\epsilon_{\rm init}$, $\epsilon_{\rm min}$, $\epsilon_{\rm decay}$, $N_r$, $N_f$, and $N_b$. \\
{\bf Initialize}: Replay buffer $\mathcal{D}$, online network $Q(s,a; \varrho)$, target network $Q^*(s,a; \varrho^*)$, $\varrho^*\leftarrow\varrho$, exploration rate $\epsilon\leftarrow\epsilon_{\rm init}$, $\boldsymbol{\rm{G}}$, and $\boldsymbol{\rm{H}}$. \\
\For{{\rm episode} $i=1, 2, ..., I$}
    {
    {\bf Reset}: Initial $s_1$ and initial $\Theta_0=0$.\\
    \For{{\rm step} $t=1, 2, ..., T$}
        {
        Use the $\epsilon$-greedy method to choose $a_t$;\\     
        Execute $a_t$ to update $\Theta_t$ based on (\ref{equi:phase});\\
        Observe $r_t$ based on (\ref{equi:RIS}) - (\ref{equi:SumRate}) and $s_{t+1}$;\\
        Store the transition \{$s_t$, $a_t$, $r_t$, $s_{t+1}$\} into $\mathcal{D}$;\\
        \If{{\rm length($\mathcal{D}$) $\geq N_b$}}
            {
            Sample a random minibatch of $N_b$ tuples \{$s_j$, $a_j$, $r_j$, $s_{j+1}$\}, $j=1, ..., N_b$ from $\mathcal{D}$;\\
            \For{$j=1, 2, ..., N_b$}
            {
                $a_{max} = \underset{a_{j+1}}{\rm arg max}~Q(s_{j+1}, a_{j+1}; \varrho)$;\\
                Calculate target values: $y_j$=
                \quad $\begin{cases}
                    r_j, & {\rm if}~done,\\
                    r_j + \gamma Q^*(s_{j+1}, a_{max}; \varrho^*), & {\rm else}.
                \end{cases}$\\
            }
            Update $\varrho$ using gradient descent with loss:\\
            \quad $L(\varrho)=\frac{1}{2N_b}\sum^{N_b}_{j=1}{(y_j-Q(s_j, a_j; \varrho))}^2$;\\
            Update $\varrho^*\leftarrow \varrho$ every $N_f$ steps;\\
            }
        }
    }

\end{algorithm}

\vspace{-5pt} 
\begin{table}[t]
\centering
\caption{Parameter setup in proposed DRL algorithms}
\vspace{-5pt} 
\footnotesize
\label{table: DRL parameter}
\begin{tabular}{|p{3.2cm}<{\centering}|p{0.9cm}<{\centering}|p{1.5cm}<{\centering}|p{1.5cm}<{\centering}|} 
\hline
Parameters & Symbols & DDQN & DDQN+GA\\
\hline
Discount factor & $\gamma$ & \multicolumn{2}{c|}{0.99} \\
\hline
Learning rate & $\alpha$ & 1e-3 & 5e-4\\
\hline
Initial exploration rate & $\epsilon_{\rm init}$ & \multicolumn{2}{c|}{1} \\
\hline
Minimum exploration rate & $\epsilon_{\rm min}$ & \multicolumn{2}{c|}{0.001} \\
\hline
Exploration decay rate & $\epsilon_{\rm decay}$ & \multicolumn{2}{c|}{0.0001} \\
\hline
Replay buffer size & $N_r$ & \multicolumn{2}{c|}{8000} \\
\hline
Minibatch size & $N_b$ & \multicolumn{2}{c|}{512} \\
\hline
Target update frequency & $N_f$ & 1000 & 2000\\
\hline
Number of episodes & $I$ & \multicolumn{2}{c|}{5000} \\
\hline
Number of steps & $T$ & \multicolumn{2}{c|}{7} \\
\hline
Reward weight of final step & $\omega$ & \multicolumn{2}{c|}{2} \\
\hline
\end{tabular}
\vspace{-10pt}
\end{table}

\subsection{Integration of Heuristic Algorithm for Element-wise Phase Shift Design}
After employing the proposed low action-space DDQN algorithm for column-wise control of RIS phase shifts, we further integrate it with a heuristic algorithm for element-wise fine-tuning. As shown in Fig.~\ref{fig:DRL} and Algorithm.~\ref{Algorithm 2}, during each DRL step, a GA is applied after the DDQN selects an action to adjust the corresponding RIS column. Specifically, the selected RIS column is first adjusted as a whole based on the DDQN action, and then further refined element by element using GA to achieve a local optimum at each DRL step. Through continuous interactions, the DDQN agent gradually learns to select actions that result in configurations more amenable to subsequent GA fine-tuning, effectively acting as an experienced guide to direct the GA’s fine-optimization. This method avoids enlarging the DRL action space while enabling detailed fine-tuning optimization, resulting in better performance than conventional column-wise controlled schemes with direct actions. For the DDQN-GA method, the DRL state $s_t$ is extended to $s_t = [s_t\ \, {\Theta'}_{t-1}]$, while the other basic settings remain the same as those of the DDQN method in Section III.A. In $s_t$ of the DDQN-GA method, ${\Theta}_{t-1}$ denotes the configuration updated solely by DDQN, and ${\Theta'}_{t-1}$ denotes the configuration further refined by GA. By including both ${\Theta}$ and ${\Theta'}$, the DNN can better learn the optimal policy, as the refinement by GA leads to a more complex mapping between actions and ${\Theta'}$, making it harder for the DNN to generalize from states. This challenge is mitigated by the stable information provided by ${\Theta}$, thereby improving the training efficiency and overall performance of the DDQN-GA algorithm.

\begin{algorithm}[t]
\caption{Integration of Greedy Algorithm}\label{Algorithm 2}
\LinesNumbered
{\bf Function}: For the proposed Heuristic-integrated DRL algorithm, {\bf Algorithm 2} is incorporated into {\bf Algorithm 1} after {\bf Line~8} by expanding states $s_t = [s_t\ \, {\Theta'}_{t-1}]$.\\
Store ${\Theta'}_t\leftarrow \Theta_t$, ${r'}_t\leftarrow r_t$;\\
Obtain vector ${\boldsymbol {\mathrm \vartheta}}$ from $\Theta_t$ according to $a_t$;\\
\For{$p=1, 2, ..., P$}{
    $p'=(p-1)~{\mathrm {mod}}~\sqrt{N}+1$;\\
    Update ${\boldsymbol {\mathrm \vartheta}}(p')$ and corresponding $\Theta_t$ based on (\ref{equi:phase});\\
    Calculate $r_t$ based on (\ref{equi:RIS}) -- (\ref{equi:SumRate});\\
    \uIf{$r_t \geq {r'}_t$}{
        ${r'}_t\leftarrow r_t$;\\
    }\Else{
        Recover ${\boldsymbol {\mathrm \vartheta}}(p')$ and corresponding $\Theta_t$;\\
    }
}
$r_t\leftarrow {r'}_t$;\\
\end{algorithm}

\vspace{-10pt}
\subsection{Complexity Analysis}
\vspace{-3pt}
In the DDQN framework, the complexity of the online network with $L$ layers is $\mathcal{O}(IT\sum_{l=1}^{L}(n_{l}n_{l-1}))$, while that of the target network is $\mathcal{O}(\frac{IT}{N_f}\sum_{l=1}^{L}(n_{l}n_{l-1}))$, where $n_{l}$ is the number of nodes in the $l$th layer. Assuming each GA process has a complexity of $\mathcal{O}(V)$, the overall embedded GA complexity is $\mathcal{O}(ITPV)$. Thus, the total complexity of the proposed algorithm is $\mathcal{O}(IT((1+\frac{1}{N_f})\sum_{l=1}^{L}(n_{l}n_{l-1})+PV))$.
\vspace{-15pt}

\section{Numerical Results}
\label{sec: result}
\vspace{-3pt}
In this section, we evaluate the performance of the proposed algorithms through numerical results and compare them against the following baselines and benchmarks:
\begin{itemize}[leftmargin=*]
    \item Flat: Set $\Theta=0$, which implies no phase-shift optimization;
    \item Random: Set $\Theta$ to randomly selected phase shifts;
    \item DQN: DQN for column-wise $\Theta$ optimization~\cite{10037180}.
    \item DDQN: DDQN~\cite{9919620} for column-wise $\Theta$ optimization.
    \item PSO: Particle swarm optimization (PSO) for continuous $\Theta$ optimization~\cite{10685472}, then quantize to discrete values.
\end{itemize}

The evaluation considers an indoor scenario modeled as a rectangular space with dimensions 8m $\times$ 8m in length and width, and a height of 6m. In this space, the rectangular RIS is fixed at the center of the ceiling, while the FBS and multiple users are randomly positioned in two separate half-spaces with heights from 0m to 4m. Their sizes are set as $N=100$, $M=4$, and $K=2$, respectively.
To represent the incident and reflect channels, we construct their LoS components based on the spatial distances between the FBS antennas, RIS elements, and users, adopting the ITU LoS basic transmission loss model~\cite{ITU}. Additionally, the NLoS components are assumed to be independently and identically distributed (i.i.d.), and the Rician factors are set as $\varepsilon_{h},\varepsilon_{G}=5$. Besides, the transmit power $P_{\rm max}=25{\rm dBm}$, the noise power $\sigma^{2}=-94{\rm dBm}$, the operating frequency $f_c=5.25{\rm GHz}$, the bandwidth $B=10{\rm MHz}$, the antennas of FBS are $1/2$ wavelength, and the unit cells of RIS are $1/4$ wavelength. For the DRL setup, the online network $Q$ and the target network $Q^*$ both consist of three fully connected layers of 512, 256, and 128 nodes for the proposed DDQN algorithm while an additional fully connected layer with 512 nodes is added at the forefront for the proposed DDQN-GA algorithm. Other training parameters are shown in Table~\ref{table: DRL parameter} for a $10\times 10$ RIS.

\begin{figure}[t]
\centering
\includegraphics[width=70mm]{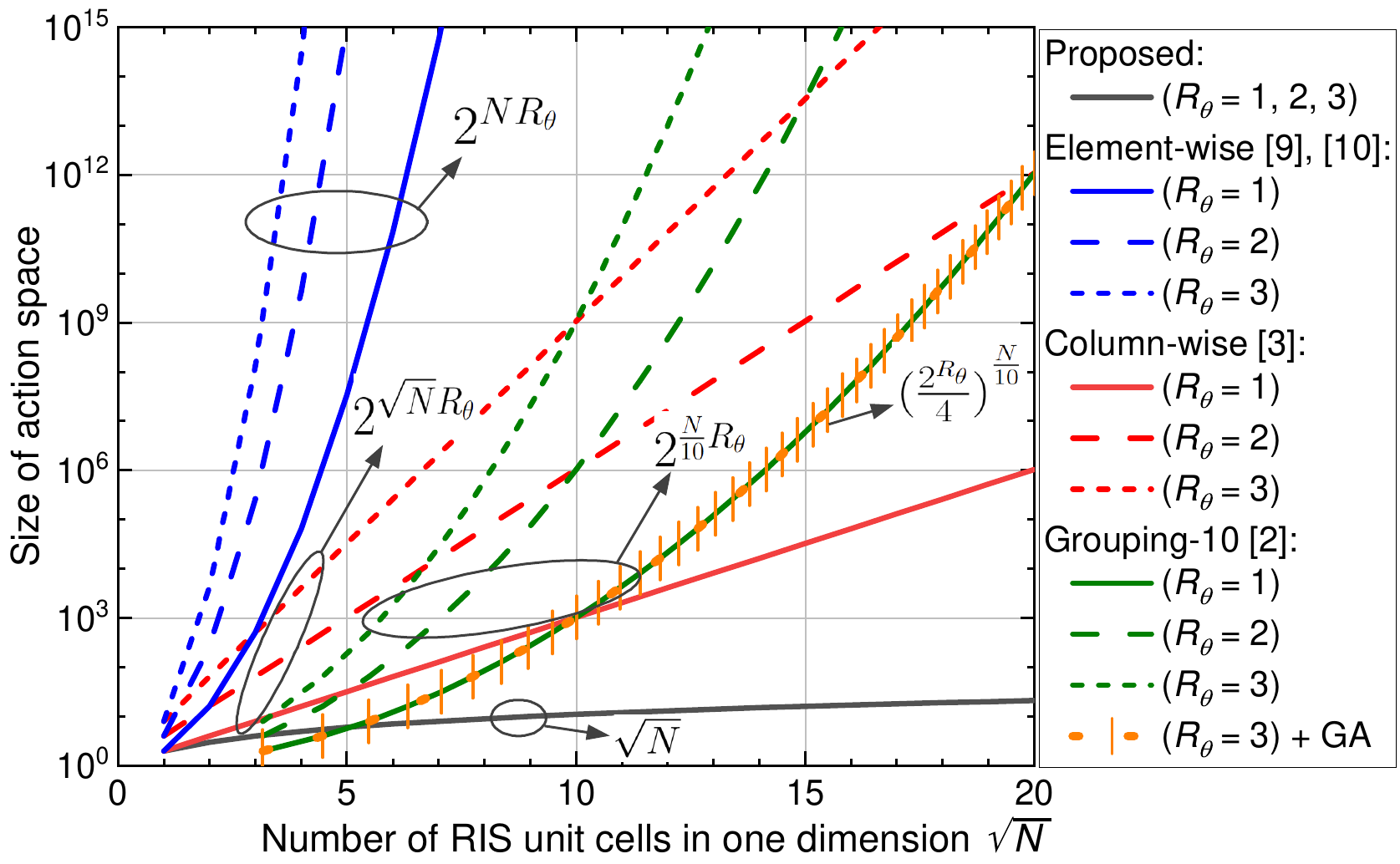}
\vspace{-5pt}
\caption{Action space size for discrete DRL algorithms with different control schemes and RIS phase shift resolutions. The x-axis represents the number of RIS unit cells in one dimension of a rectangular RIS array.}
\label{fig:curve}
\vspace{-10pt}
\end{figure}

Fig.~\ref{fig:curve} displays the size of the discrete DRL action space for different action setups. As shown in this figure, applying element-wise control for RISs~\cite{9277627, 9526285} results in the fastest growth of action space size as the number of RIS unit cells increases. For a $5\times 5$ RIS, their action space sizes for 1-bit, 2-bit, and 3-bit resolutions are $2^{1\times 25}\approx 10^{7}$, $2^{2\times 25}\approx 10^{15}$, and $2^{3\times 25}\approx 10^{22}$, respectively, which are difficult (if not impossible) for discrete DRL to optimize and converge~\cite{10021676}. By grouping 10 adjacent cells~\cite{10500737} or unit cells in each column~\cite{10037180} to share the same phase shift, element-wise control is simplified to column/group-wise control, significantly reducing the action space size. However, when expanding to $20\times 20$ RISs, the action space sizes for column-wise control still reach $2^{1\times 20}\approx 10^{6}$, $2^{2\times 20}\approx 10^{12}$, and $2^{3\times 20}\approx 10^{18}$ for 1-, 2-, and 3-bit resolutions, respectively, let alone offering a lower phase shift DoF than element-wise control. Compared to the foregoing approaches, the proposed method, which accumulates the actions over multiple steps to represent the phase shift configuration, still maintains a low action space when optimizing significantly larger RISs, such as $20\times20$ or even $50\times50$ arrays with acceptable action space sizes between $10^1-10^2$. Meanwhile, the integration of the greedy algorithm also maintains an element-wise DoF in adjusting the RIS phase shifts to avoid the performance degradation of grouping. These features provide the proposed algorithms with superior adaptability for optimizing large-scale RISs.

\begin{figure*}[t]
\centering 
    \subfloat[\label{fig:StepPerEpisode}]{
      \begin{minipage}[t]{0.3\linewidth}
        \includegraphics[width=0.92\columnwidth]{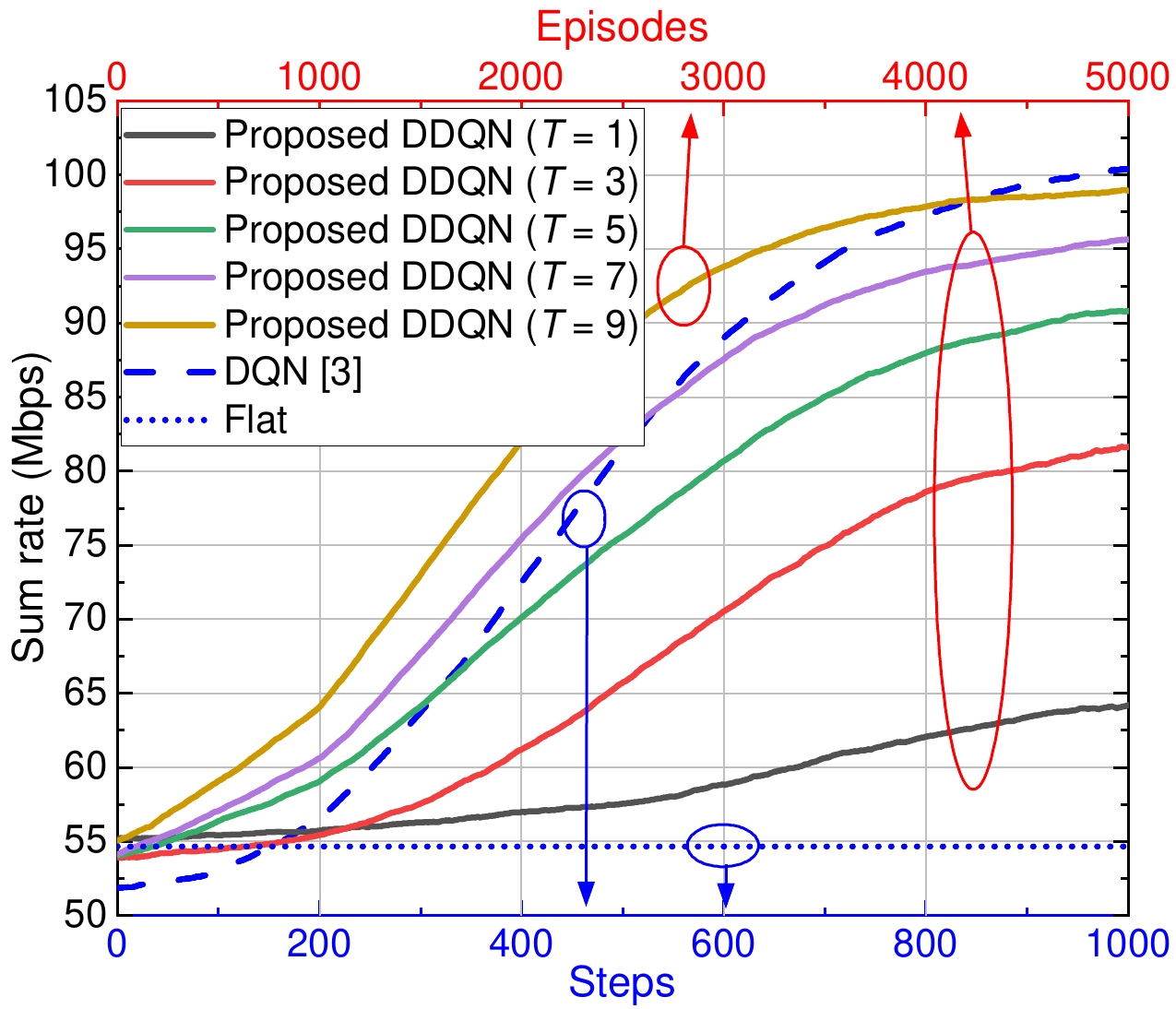}   
      \end{minipage}%
      }
    \subfloat[\label{fig:Training}]{
      \begin{minipage}[t]{0.3\linewidth}   
        \includegraphics[width=0.92\columnwidth]{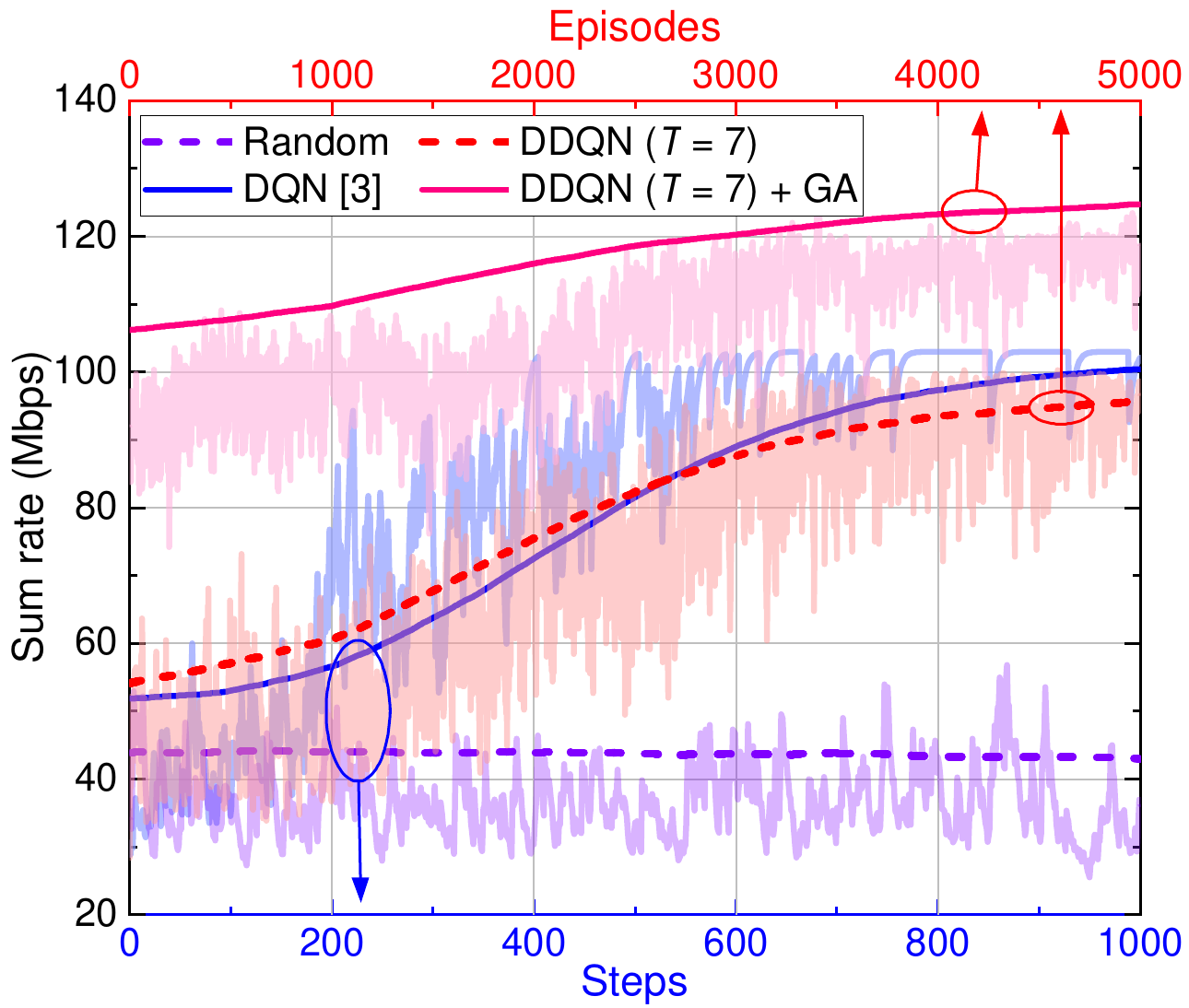}   
      \end{minipage} 
      }
    \subfloat[\label{fig:CompareCrossMethods}]{
      \begin{minipage}[t]{0.29\linewidth}   
        \includegraphics[width=0.92\columnwidth]{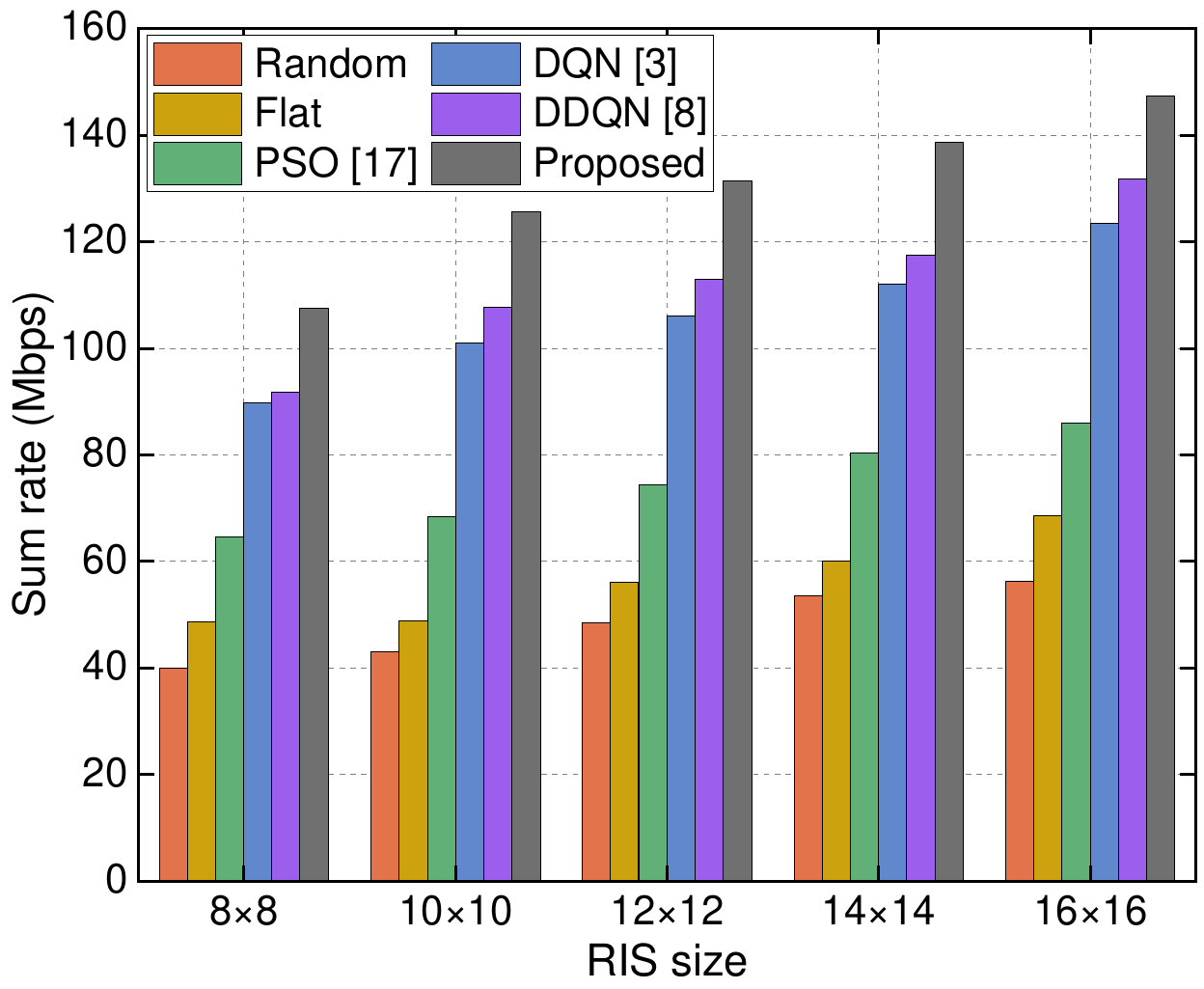}
      \end{minipage} 
      }
\caption{Numerical results: (a) Averaged and smoothed DRL training curves under different numbers of steps per episode; (b) Instant and averaged DRL training curves for various optimization methods; (c) Sum rate versus RIS size.} \label{fig:Result}
\vspace{-10pt}
\end{figure*}

Since the accumulated adjustments over multiple steps within each episode are used to represent a possible RIS phase-shift configuration, as shown in Fig.~\ref{fig:DRL}, we investigate the impact of the number of steps per episode on optimization performance. As illustrated in Fig.~\ref{fig:StepPerEpisode}, for a $10 \times 10$ RIS with 1-bit resolution, the proposed DDQN algorithm achieves improved optimization performance as the number of steps per episode increases, and it achieves a comparable sum rate to the benchmark when $T$ is around 9. These results indicate that the proposed DRL approach, which significantly reduces the large action space to a much smaller one, also effectively optimizes RIS phase shift configurations, albeit at the cost of increased convergence iterations.

After validating the effectiveness and performance of the proposed low action-space DRL approach in column-wise RIS phase-shift optimization, we further investigate the integration of the GA for element-wise optimization. 
In Fig.~\ref{fig:Training}, the proposed algorithms are compared against the Random and DQN benchmarks based on a $10 \times 10$ RIS. The light-colored lines represent a set of instant training curves, while the dark-colored lines represent the smoothed average curves across multiple randomly generated scenario setups.
The proposed DDQN algorithm is set to $T=7$ to balance training complexity and performance. Although its performance is $5\%$ lower than DQN, integrating GA optimization at each DRL step enables the DDQN-GA algorithm to achieve a $30\%$ higher sum rate than DDQN without increasing the convergence steps and $24\%$ higher than DQN.

Fig.~\ref{fig:CompareCrossMethods} illustrates the comparison of the proposed DDQN-GA algorithm with other benchmark schemes for different RIS sizes. Although the element-wise controlled PSO scheme achieves better performance in the continuous phase-shift space~\cite{10685472}, quantization for discrete spaces leads to significant performance degradation, especially for low-resolution RIS like 1-bit, limiting its effectiveness in optimizing discrete RISs. Compared to DQN~\cite{10037180} and DDQN~\cite{9919620} in optimizing column-wise controlled RISs, the proposed DDQN-GA achieves significantly better performance by adopting the proposed accumulated action method and integrating GA, although it may require slightly longer training time. However, this does not affect the efficiency of DRL in achieving high-quality discrete RIS configurations quickly during inference.

\section{Conclusion}
\label{sec: conclusion}
In this letter, we propose a novel low-action-space DDQN-GA algorithm to optimize the phase shift configurations of RISs. Unlike most existing literature that defines the entire set of potential RIS configurations as the DRL action space, our approach represents a potential RIS configuration using the accumulated incremental phase over multiple steps. Furthermore, we integrate this method with the GA at each step to achieve finer optimization. This proposed algorithm effectively reduces the DRL action space, enabling its application in optimizing large-scale RISs.
\vspace{-5pt}
\bibliographystyle{IEEEtran}
\bibliography{IEEEabrv,references} 

\end{document}